\documentclass[twocolumn,showpacs,preprintnumbers]{revtex4}
\usepackage{amsfonts}
\usepackage{amssymb}
\usepackage{amsmath}
\usepackage{graphicx}
\usepackage{dcolumn}
\usepackage{bm}
\usepackage{color}

\begin{document}

\title{Superradiant phase transitions in the  quantum Rabi model: Overcoming the no-go theorem through anisotropy	}
\author{Tian Ye$^{1}$ }
\author{Yan-Zhi Wang$^{1}$}
\author{Xiang-You Chen$^{2,3}$}
\author{Qing-Hu Chen$^{3,4,}$}
\email{qhchen@zju.edu.cn}
\author{Hai-Qing Lin$^{5,}$}
\email{hqlin@zju.edu.cn }

\affiliation{$^{1}$Anhui Province Key Laboratory for Control and Applications of Optoelectronic Information Materials, Department of Physics, Anhui Normal University, Wuhu, Anhui 241000, China\\
$^{2}$School of Physics and Astronomy, Yunnan University, Kunming, Yunnan 650091, China\\
$^{3}$Zhejiang Key Laboratory of Micro-Nano Quantum Chips and Quantum Control, School of Physics, Zhejiang University, Hangzhou Zhejiang 310027, China\\
$^{4}$Collaborative Innovation Center of Advanced Microstructures, Nanjing University, Nanjing, Jiangsu 210093, China\\
$^{5}$Institute for Advanced Study in Physics and School of Physics, Zhejiang University, Hangzhou, Zhejiang 310058, China}

\date{\today }

\begin{abstract}
Although the superradiant phase transition (SRPT) is prohibited in the paradigmatic quantum Rabi model due to the no-go theorem caused by the $\mathbf{A}^2$ term, we demonstrate two distinct types of SRPTs emerging from the normal phase in the anisotropic quantum Rabi model.	A discontinuous phase transition between the two types of superradiant phases also emerges in the presence of a strong $\mathbf{A}^2$ term. Additionally, a rich phase diagram featuring a triple point, which connects first- and second-order phase transitions, is derived analytically and confirmed through numerical diagonalization at large effective system sizes.	Finally, distinct critical behavior at the triple point is revealed and contrasted with that of a single continuous SRPT.	This work may open a new avenue for observing SRPTs in their intrinsic form without altering the $\mathbf{A}^2$ term, while also offering a practical platform for exploring rich quantum phenomena.	
\end{abstract}

\pacs{05.30.Rt,
42.50.Ct,
42.50.Pq,
05.70.Jk}
\maketitle

\section{\label{Introduction} Introduction }

The Dicke model (DM) describes the interaction between a collection of two-level atoms (qubits) and a single-mode photon field \cite{DM}, making it a paradigmatic model in quantum optics.	In the thermodynamic limit of infinite qubits, the DM theoretically undergoes a superradiant phase transition (SRPT), characterized by macroscopic excitation of the photon field, at both finite \cite{DM_PT, DM_PT2} and zero temperatures \cite{emary2003prl}.	

Observing an equilibrium SRPT in experiments has long been challenging due to the inability to access sufficiently strong qubit-cavity coupling in cavity quantum electrodynamics (QED).	Recent advancements in solid-state quantum platforms and quantum simulations, such as superconducting circuit QED \cite{tniemczyk2010np,forn2010prl,yoshihara2016np}, trapped ions \cite{trap_ions,lmduan2021nc}, and cold atoms \cite{cold_atom}, have enabled light-matter interactions to reach the ultrastrong- and even deep strong-coupling regimes.	It has inspired several theoretical works on the SRPT in the DM \cite{cuiti2014prl,keeling2019aqt}.	In particular, the few-component SRPT has been explored in the quantum Rabi model (QRM) \cite{hwang2015prl}, which is a single-atom version of the DM \cite{QRM}, in the limit of an infinite qubit-to-cavity frequency ratio. {The few-component SRPT exhibits the same characteristic of cavity-field macroscopic excitation as the DM~\cite{hwang2015prl}}. Since then, the SRPT has been studied in various few-component models, including the Jaynes-Cummings model \cite{hwang2016prl}, the anisotropic QRM \cite{mxliu2017prl,shen2017pra}, and the quantum Rabi triangle \cite{zhang2021prl}. Notably, the SRPTs of the DM and QRM belong to the same universality class, even with anisotropic qubit-cavity coupling \cite{mxliu2017prl}.	

Despite this renewed possibility, the experimental realization of the SRPT remains debated and an open question to this day~\cite{A2original,keeling2007jpb,vukics2014prl,ciuti2010nc,viehmann2011prl,ciuti2012prl, viehmann_arxiv,rabl2016pra,bamba2016prl,bamba2017pra,andolina2019prb,nori2019np,nazir2020prl,nazir2022rmp}.	In cavity QED, Rza\.{z}eski \emph{et al.} argued that the $\mathbf{A}^2$ term, originating from the minimal coupling Hamiltonian, would forbid the SRPT of the DM at any finite coupling strength if the Thomas-Reich-Kuhn (TRK) sum rule for the atom is properly considered~\cite{A2original}. {In detail, the TRK sum rule dictates that the sum of oscillator strengths for all electronic dipole transitions to any energy level in an atom equals unity. It then establishes a minimum Dicke-coupling-dependent strength of the $\mathbf{A}^2$ term, which prevents the SRPT of the DM.} This no-go theorem also applies to the SRPT of the QRM \cite{hwang2015prl}. Based on the effective model of circuit QED, Nataf and Ciuti proposed that the TRK sum rule in cavity QED could be violated, allowing the SRPT to occur in principle \cite{ciuti2010nc}.	 Viehmann \emph{et al.} questioned whether, even with a complete microscopic treatment, the no-go theorem of cavity QED still applies to circuit QED \cite{viehmann2011prl}.	These arguments focus on whether the TRK sum rule differs in various systems \cite{viehmann2011prl,ciuti2012prl}, raising a significant controversy over whether the $\mathbf{A}^2$ term can be engineered in the superconducting qubit circuit platform \cite{viehmann_arxiv,rabl2016pra, bamba2016prl,andolina2019prb}.	On the other hand, the validity of the two-level approximation, a crucial step in constructing the DM in various solid-state devices,and the applicability of Coulomb and electric dipole gauges and different experimental schemes leading to the varying $\mathbf{A}^2$ term, are widely discussed in the literature (see \cite{nori2019np,nazir2020prl,nazir2022rmp,Andolina}, and references therein).	 A consensus has yet to be reached, and the no-go theorem remains the most controversial subject.	

To overcome the no-go theorem and realize the SRPT, several groups have introduced additional extrinsic factors that extend beyond the original DM and QRM.	It was suggested that the no-go theorem does not apply to the three-level DM, making the SRPT possible \cite{ciuti2013pra}.	By adding an interaction between atoms in the DM, Liu \emph{et al.}  found that an SRPT can occur in a specific parameter regime \cite{taoliu2012pla}.	Jaako \emph{et al.} proposed that a circuit QED system with a symmetrically coupled single resonator can contain interactions between qubits, enabling the achievement of an SRPT in the two-qubit QRM \cite{rabl2016pra}.	Considering the hopping between cavity bosons at two sites, the competition between the $\mathbf{A}^2$ term and the boson hopping term can lead to the appearance of the SRPT in the two-site QRM \cite{ymwang2020pra}.	Recently, Chen \emph{et al.} introduced an antisqueezing effect into the NMR quantum simulator, directly alleviating the constraint of the no-go theorem and realizing the SRPT in the QRM with the $\mathbf{A}^2$ term \cite{Peng}.	

It would be interesting to go beyond the no-go theorem intrinsically by manipulating the internal factors of the original DM and QRM without altering the $\mathbf{A}^2$ term subject to the TRK sum rule.	In this paper, we investigate the anisotropic QRM with varying coupling strengths of the rotating-wave and counter-rotating-wave terms and arbitrary $\mathbf{A}^2$ term.	It is well known that the anisotropic QRM without the $\mathbf{A}^2$ term can exhibit an SRPT~\cite{mxliu2017prl,shen2017pra}.	Recently, multiple ground-state instabilities have been observed in the anisotropic QRM \cite{xychen}, driven solely by the anisotropy.	A natural question arises: Can the SRPT be induced by anisotropy in the anisotropic QRM with the $\mathbf{A}^2$ term?	Interestingly, we find not only two types of SRPT, but also first-order phase transitions between them, and further construct a rich phase diagram that includes a triple point.	

The paper is organized as follows: In Sec.~\ref{sec2}, we briefly introduce the anisotropic quantum Rabi model with the $\mathbf{A}^2$ term.	In Sec.~\ref{sec3}, using the Bogoliubov transformation, we derive a renormalized anisotropic QRM without the $\mathbf{A}^2$ term.	We then analytically obtain two types of SRPT and a rich phase diagram in the large limit of the frequency ratio.	These analytical findings are confirmed numerically in Sec.~\ref{sec4}, where the distinguishing critical behavior at the triple point is also detected through finite-size scaling analysis. A summary is given and conclusions are drawn in Sec.~\ref{conclusion}.	The Appendix analytically presents the energy-gap critical exponents at both the triple point and the standard SRPT based on an effective Hamiltonian.	

\section{\label{model} The anisotropic QRM with the $\mathbf{A}^2$ term \label{sec2}}

The anisotropic QRM with the $\mathbf{A}^2$ term can be described as follows.	
\begin{eqnarray}
H &=&\omega a^{\dagger }a+\frac{\Delta }{2}\sigma _{z}+g\left[ (a\sigma
_{+}+a^{\dagger }\sigma _{-})+\tau(a\sigma _{-}+a^{\dagger }\sigma _{+})\right]
\notag \\
&+&D(a+a^{\dag })^{2},
\end{eqnarray}%
where the  operator $a^{\dagger}$ ($a$) creates (annihilates) a photon of the electromagnetic mode with frequency $\omega$, while $\Delta$ represents the qubit energy splitting. The operators $\sigma_{\pm} = \frac{1}{2}(\sigma_x \pm i \sigma_y)$ excite (relax) the qubit from $\left\vert + \right\rangle$ ($\left\vert - \right\rangle$) to $\left\vert - \right\rangle$ ($\left\vert + \right\rangle$), where $\left\vert \pm \right\rangle$ denotes the excited or ground state of the qubit. Here, $\sigma_i$ ($i=x,y,z$) represents the Pauli matrices. The coupling strength is denoted by $g$, $\tau$ is the anisotropic parameter of the coupling (which can take both positive and negative values for various applications), and $D$ is the strength of the $\mathbf{A}^2$ term.

The anisotropic QRM bridges the gap between the QRM and the Jaynes-Cummings model. Specifically, as the anisotropic parameter varies from $\tau=1$ to $\tau=0$, the model transitions continuously from the QRM to the Jaynes-Cummings model. Notably, the anisotropic QRM retains the same $\mathbb{Z}_2$ symmetry as the QRM, with the parity operator $P = -\sigma_z e^{i \pi a^\dagger a}$, even when the $\mathbf{A}^2$ term is included.	

To relate to the no-go theorem for the SRPT in the QRM, we define the strength of the $\mathbf{A}^2$ term as $D = \kappa D_0$, where $\kappa$ is a nonnegative constant and $D_0 = \frac{g^2}{\Delta}$. When $\kappa = 1$, it represents the minimum strength of the $\mathbf{A}^2$ term, as dictated by the TRK sum rule for the atom.	In the original light-matter interaction platform, where a real atom is coupled to the photonic mode, the $\mathbf{A}^2$ term arises from $(\bm{p} - q \bm{A})^2 / 2m$ in the minimal coupling Hamiltonian, where $\bm{p}$ and $\bm{A}$ represent the electron momentum operator and the electromagnetic vector potential, respectively.	Theoretically, this $\mathbf{A}^2$ term is often neglected in the literature, even in the strong coupling regime, which leads to the SRPT being present and extensively studied for $\kappa = 0$.	Since the $\mathbf{A}^2$ term may be non-negligible in both cavity and circuit QEDs in various ways, its strength can be adjusted to arbitrary values for broader applications.	

Due to the ongoing progress in quantum technology, the anisotropic QRM can now be realized using various schemes, such as superconducting qubit circuits \cite{qtxie2014prx, cuiti2014prl, wjyang2017pra}, electron gases with both Rashba and Dresselhaus spin-orbit interactions \cite{erlingsson2010pra}, and spin qubits coupled to anisotropic ferromagnets \cite{skogvoll2021pra}. Specifically, in circuit QED, a large (superconducting quantum interference device) SQUID generates an electromagnetic field and is inductively coupled to a qubit realized by another SQUID. By including both the circuit inductance and the mutual inductance between the two SQUIDs, the anisotropic qubit-photon coupling can be simulated, provided that the capacitive interaction between these two SQUIDs is engineered to be negligible.   {Another realization of the anisotropic QRM has been proposed using an ensemble of nitrogen-vacancy center spins in diamond, coupled to the quantized magnetic field of a superconducting microwave cavity~\cite{ZouPRL2014}.	A driving field induces coupling between two excited states of an artificial atom through two-photon Raman transitions within the cavity.	The two possible transitions between the atomic states involve either the absorption or emission of a cavity photon. This results in both rotating-wave terms $a^{\dagger}\sigma_{-}$ and $a\sigma_{+}$ and counter-rotating-wave terms $a^{\dagger}\sigma_{+}$ and $a\sigma_{-}$, thereby yielding the anisotropic QRM. In these experiments, the effect of the $\mathbf{A}^2$ term is worth considering.

\section{Recovery of the Superradiant Phase Transition through Anisotropy \label{sec3}}

\subsection{Bogoliubov transformed Hamiltonian\label{sec3A}}
To facilitate the analytical study, we first eliminate the $\mathbf{A}^2$ term by introducing new bosonic operators, derived from the Bogoliubov transformation applied to the original operators.
\begin{subequations}
\begin{align}
b& =\mu a+\nu a^{\dagger }, \\
b^{\dagger }& =\mu a^{\dagger }+\nu a.
\end{align}%
To ensure the bosonic nature of the new operators, the coefficients must satisfy the constraint $\mu^2 - \nu^2 = 1$. If the condition $D(\mu - \nu)^2 - \omega \mu \nu = 0$ is satisfied, the original Hamiltonian can be transformed into a form without the $\mathbf{A}^2$ term,
\end{subequations}
\begin{equation}
H=\omega ^{\prime }b^{\dagger }b+\frac{\Delta }{2}\sigma _{z}+g^{\prime }%
\left[ (b\sigma _{+}+b^{\dagger }\sigma _{-})+\tau^{\prime }(b\sigma
_{-}+b^{\dagger }\sigma _{+})\right] ,  \label{Hbf}
\end{equation}%
where $\omega^{\prime } = (\mu + \nu)^2 \omega$, $g^{\prime } = g(\mu - \tau \nu)$, and $\tau^{\prime } = (\tau \mu - \nu) / (\mu - \tau \nu)$. For later reference, we outline two constraint conditions for $\mu$ and $\nu$,
\begin{eqnarray}
\gamma &=&(\mu +\nu )^{2}=\sqrt{1+\kappa \tilde{g}^{2}},  \label{munu_sum} \\
\zeta &=&\frac{\nu }{\mu }=1+\frac{2}{\kappa \tilde{g}^{2}}-\sqrt{(1+\frac{2%
}{\kappa \tilde{g}^{2}})^{2}-1}.  \label{munu_rate}
\end{eqnarray}%
Here, the rescaled coupling strength is given by $\tilde{g} = 2g / \sqrt{\omega \Delta}$. It is important to note that the new transformed parameters $g^\prime$ and $\tau^\prime$ can extend into the negative regime, even when the original parameters $g > 0$ and $\tau > 0$. This extension may lead to a richer phase diagram than that of the standard anisotropic QRM.

\subsection{Superradiant phase transitions enriched by anisotropy \label {sec3B}}

In the standard QRM, the effective system size is defined by the frequency ratio $\eta = \Delta / \omega$. The quantum phase transition (QPT) occurs in the thermodynamic limit, i.e., as $\eta \to \infty$, while the rescaled coupling strength $\tilde{g}$ remains finite. To investigate the SRPT in the current anisotropic model with the $\mathbf{A}^2$ term, we similarly define $\eta' = \Delta/\omega^{\prime} \to \infty$ and keep the auxiliary coupling $\tilde{g}^{\prime} = 2g^{\prime}/\sqrt{(\omega^{\prime} \Delta)}$ finite.

In terms of the field quadratures, $x = (b + b^\dagger)/\sqrt{2}$ and $p =i(b^\dagger - b)/\sqrt{2}$, we rewrite the Hamiltonian (\ref{Hbf}) in the following rescaled form:
\begin{eqnarray}  \label{Hrscl}
\tilde{H} &=&\frac{H}{\Delta }=\frac{x^{2}+p^{2}}{2\eta ^{\prime }}+\frac{1}{%
2}\sigma _{z}  \notag \\
&+&\frac{\tilde{g}^{\prime }}{\sqrt{8\eta^{\prime }}}\left[ (1+\tau^{\prime
})x\sigma _{x}-(1-\tau^{\prime })p\sigma _{y}\right] .
\end{eqnarray}%
In the thermodynamic limit, the coupling term becomes the leading-order perturbation, proportional to $1/\sqrt{\eta^{\prime}}$. In particular, this term is off-diagonal in the basis of the eigenvectors of $\sigma_z$. To diagonalize this perturbation, we perform the unitary transformation $H_{\text{eff}}=e^{-S}\tilde{H}e^{S}$, where the $S=-i\tilde{g}
^{\prime }\left[(1+\tau^{\prime })x\sigma_{y}+(1-\tau^{\prime })p\sigma _{x}\right]%
/\sqrt{8\eta ^{\prime }} $. {Retaining terms up to the second-order perturbation in $1/\sqrt{\eta}$, we obtain}
\begin{equation*}
H_{\text{eff}}\ \simeq \frac{\gamma (x^{2}+p^{2})}{2\eta }+\frac{1}{2}\sigma
_{z}+\frac{\gamma \tilde{g}^{2}}{8\eta }\left[ \xi
_{x}^{2}x^{2}+(1-\tau)^{2}p^{2}\right] \sigma _{z},
\end{equation*}%
where the coefficient is given by $\xi _{x}=(1+\tau)(1-\zeta )/(1+\zeta )$.

Focusing on the low-energy states, we project the Hamiltonian onto the spin-down subspace, yielding the effective low-energy Hamiltonian
\begin{equation}
H_{\text{eff}}\ \simeq \frac{\gamma }{8\eta }\left\{ 4\left(
x^{2}+p^{2}\right) -\tilde{g}^{2}\left[ \xi _{x}^{2}x^{2}+\left( 1-\tau\right)
^{2}p^{2}\right] \right\} .  \label{Heff}
\end{equation}%
Obviously, the first term dominates at weak coupling, indicating the system behaves as a normal oscillator. In the ground state, this corresponds to the normal phase, characterized by the vacuum photonic state. On the other hand, if $\tilde{g}$  becomes large enough, {the ground-state energy of this effective Hamiltonian  would  become unbounded from below due to the dominant negative coupling term. The competition between these two terms results in a phase transition}. Specifically, this leads to the emergence of $x$-type ($p$-type) superradiant phases (SRP), signaled by the divergent value of  $\langle x^{2}\rangle $ ($\langle p^{2}\rangle $)   in the ground state, similar to what was observed in the original anisotropic QRM. \cite{mxliu2017prl}

Equating the coefficients of  $x^{2}\ $\ and $p^{2}$ in the effective Hamiltonian (\ref{Heff}) gives
\begin{equation}
\kappa _{c}=4\tau/(1-\tau)^{2}\tilde{g}^{2},  \label{kappa_c}
\end{equation}%
{where $\tilde{g}$ and $\tau$ are the rescaled strength and anisotropic parameter of the coupling, and $\kappa$  represents the strength of the $\mathbf{A}^2$ term, with $\kappa=1$ corresponding to the minimum $\mathbf{A}^2$-term strength as dictated by the TRK sum rule.} The $p^{2}$-type ($x^{2}$-type) term dominates when  $\kappa >\kappa _{c}$ ($%
\kappa <\kappa_{c}$) from the perspective of macroscopic photon mode excitation. When $\kappa >\kappa _{c}$, the vanishing coefficient of the $\ p^{2}$ -type term in the effective Hamiltonian (\ref{Heff}) determines the critical coupling point that separates the $p$-type SRP from the normal phase,
\begin{equation}
\tilde{g}_{c}^{p}=\frac{2}{\left\vert 1-\tau\right\vert }  \label{gp_c}
\end{equation}%
at $\tau\neq 1$ and $\tau<\kappa $ obtained by $\kappa >\kappa _{c}(\tau,\tilde{g}%
_{c}^{p})$ . Similarly, if $\kappa <\kappa _{c}$, the vanishing coefficient of the $x^{2}$-type term yields the critical coupling strength that separates the $x$-type SRP from the normal phase,
\begin{equation}
\tilde{g}_{c}^{x}=\frac{2}{\sqrt{(1+\tau)^{2}-4\kappa }},  \label{gx_c}
\end{equation}%
at $\tau>\kappa $ obtained by $\kappa <\kappa _{c}(\tau,\tilde{g}_{c}^{x})$.

Equations~(\ref{gp_c}) and (\ref{gx_c}) describe the critical coupling of the anisotropic QRM without the $\mathbf{A}^2$ term ($\kappa = 0$) \cite{mxliu2017prl,shen2017pra}.	These equations are also consistent with the no-go theorem, which states that for $\kappa \ge 1$ and $\tau=1$, $\tilde{g}_c^p$ tends to infinity, and a real solution for $\tilde{g}_c^x$ is only possible when $\kappa < 1$ and $\tau = 1$.	

Note that $\tilde{g}_c^p$ and $\tilde{g}_c^x$ are subject to $\kappa > \kappa_c$ and $\kappa < \kappa_c$, respectively.	From Eq. (\ref{kappa_c}), $\kappa_c$ depends on the rescaled coupling strength $\tilde{g}$.	For a given $\kappa $, when the coupling constant varies and
crosses a critical value $\tilde{g}_{c}^{1}=2\sqrt{\tau/\kappa }/{|1-\tau|}$ corresponding to $\kappa =\kappa _{c}$, the macroscopically excited photon field for $x^{2}$ type to $p^{2}$ type are interchanged. This implies a first-order phase transition from the $x$-type SRP to the $p$-type SRP at the coupling $\tilde{g}_c^1$, with $\tau > \kappa$.	Interestingly, $\tilde{g}_{c}^{p}=$ $\tilde{g}%
_{c}^{x}$ would give a triple critical point at $\tau=\kappa >1$,
\begin{equation}
\tilde{g}_{c}^{\text{tri}}=\frac{2}{\tau-1}.  \label{gtri}
\end{equation}%
A triple critical point does not exist in the anisotropic QRM without the $\mathbf{A}^2$ term.	
\begin{figure}[tbp]
\centering
\includegraphics[width=\linewidth]{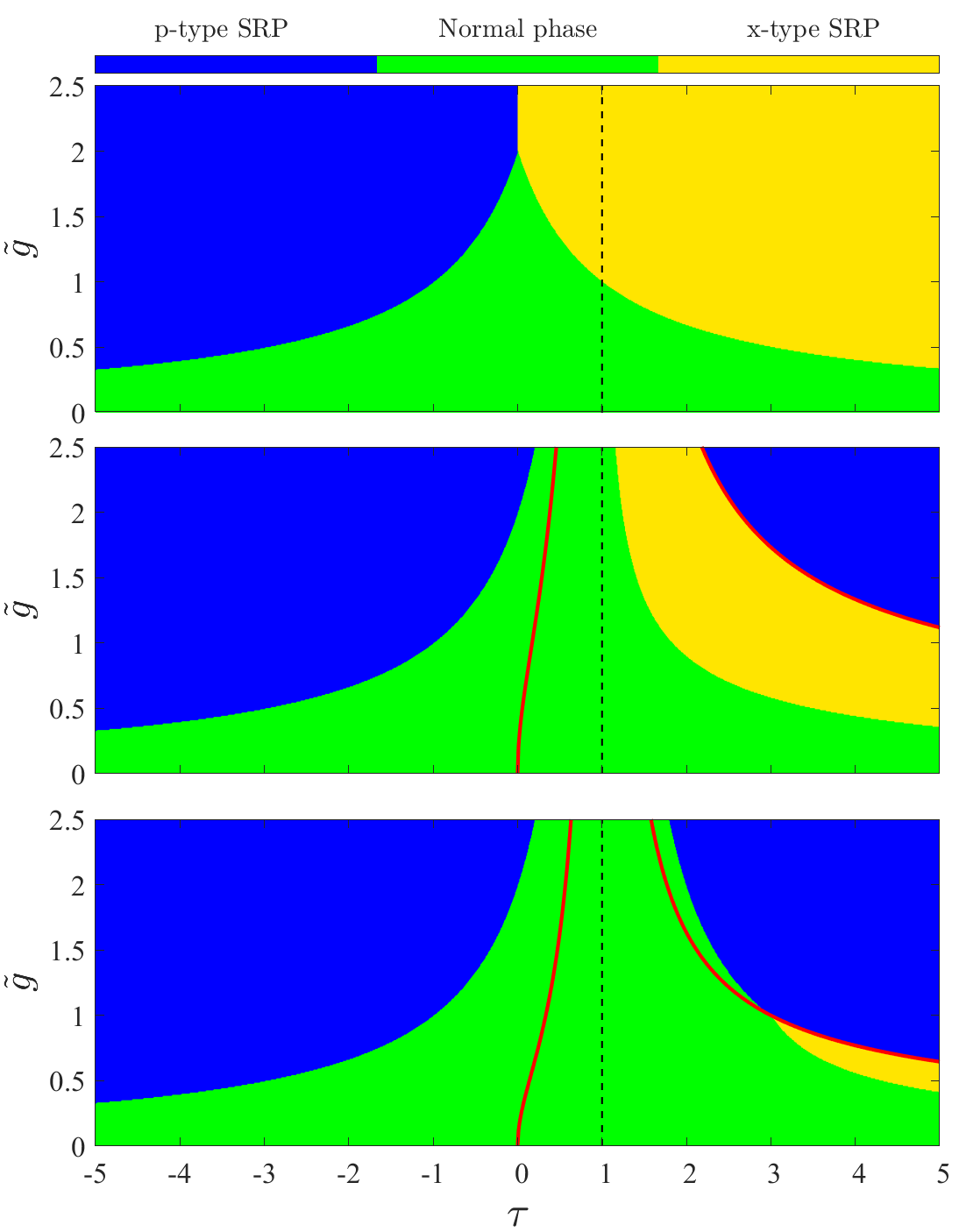}
\caption{Phase diagrams of the anisotropic QRM with three typical $\mathbf{A}^2$ terms: (a) $\kappa = 0$, (b) $\kappa = 1$ (minimum value constrained by the TRK sum rule), and (c) $\kappa = 3$. The $p$- and $x$-type SRPs are indicated in blue and yellow, respectively, while the normal phase is shown in green. Red solid lines represent $\kappa = \kappa_c$, as given by Eq. (\ref{kappa_c}).	}
\label{fig_PD}
\end{figure}
The main analytical results are shown in Fig.~\ref{fig_PD}.	Using Eqs.~(\ref{gp_c}) and (\ref{gx_c}), we plot the boundary of the $x$-type (or $p$-type) SRP for three typical values of the $\mathbf{A}^2$ strength: $\kappa = 0$ [Fig.~\ref{fig_PD}(a)], $\kappa = 1$ [Fig.~\ref{fig_PD}(b)], and $\kappa = 3$ [Fig.~\ref{fig_PD}(c)].	 The red solid curves $\tilde{g}=2\sqrt{\tau/\kappa }/|1-\tau|$ corresponding to $\kappa =\kappa _{c}$ from Eq.~(\ref{kappa_c}) are also plotted in Fig.~\ref{fig_PD}(b) and ~\ref{fig_PD}(c).

As shown in Fig.~\ref{fig_PD}(a) for $\kappa = 0$, Eqs.(\ref{gp_c}) and ~(\ref{gx_c}) recover the $x$-type and $p$-type SRPs of the anisotropic QRM without the $\mathbf{A}^2$ term \cite{mxliu2017prl}.	The phase diagram is symmetric, with $x$-type and $p$-type SRPs emerging for positive and negative $\tau$, respectively. One type of SRP transitions to the other as $\tau$ approaches zero \cite{mxliu2017prl}.	

This symmetric structure is broken in the presence of the $\mathbf{A}^2$ term, as shown in (b) and (c).	In the positive-$\tau$ regime, the $x$-type SRP shrinks and is gradually replaced by the $p$-type SRP as the $\mathbf{A}^2$ term increases.	For any finite $\kappa > 0$, Eq.~(\ref{gp_c}) shows that the $p$-type SRPT of the anisotropic QRM with the $\mathbf{A}^2$ term can always be realized.	In the case of $\tau > 0$, as long as $\tau > \kappa$, the x-type SRPT can also be realized.	\textsl {In other words, for any $\mathbf{A}^2$ strength $\kappa$, the SRPT can be driven by anisotropy as long as $\tau \neq 1$, thus overcoming the no-go theorem.	}

The analytical results above highlight a remarkable feature of the SRPT induced by anisotropy.	The p-type SRPT described by Eq.~(\ref{gp_c}) is independent of the $\mathbf{A}^2$ term as long as $\tau < \kappa$.	The effective Hamiltonian~(\ref{Heff}) explicitly demonstrates the robustness of the p-type SRPT. Note that the relative strength of two $p^{2}$-type terms is
unaffected by the $\mathbf{A}^2$ term, indicating that {the ground-state energy of the effective Hamiltonian
(\ref{Heff}) must become unbounded from below} if $\tilde{g}>\frac{2}{|1-\tau|}$.
Furthermore, {the unbound ground-state energy} implies macroscopic excitation of the photon mode and the eventual emergence of the SRP}.	Thus, combined with Eq.~(\ref{gx_c}), it is clear that the x-type SRPT is suppressed as the $\mathbf{A}^2$ term increases, while the p-type SRPT remains robust even with arbitrarily strong $\mathbf{A}^2$ terms.	At $\tau = 1$, the p-type SRPT is forbidden regardless of the $\mathbf{A}^2$ term, due to the vanishing $p^2$-type coupling ($\propto (1 - \tau)^2$) in Hamiltonian~(\ref{Heff}).

Recently, the SRPT in the Dicke-Stark model with $\mathbf{A}^2$ terms was studied  at both zero and finite temperatures~\cite{Dicke_Stark}.	Using the Holstein-Primakoff transformation of the angular momentum operator, the critical coupling strength for the {SRPT} is derived. The analytical derivation reveals, somewhat ambiguously, that the phase transition occurs for the anisotropy constant $\tau > 1$, even in the absence of the Stark coupling terms. In this work, we find that the {SRPT} can occur in both  the $\tau > 1$ and $\tau < 1$ regions, revealing a much richer phase diagram in the context of the QRM.

\section{Numerical verification and finite-frequency scaling analysis	 \label%
{sec4}}

\begin{figure}[tbp]
\centering
\includegraphics[width=\linewidth]{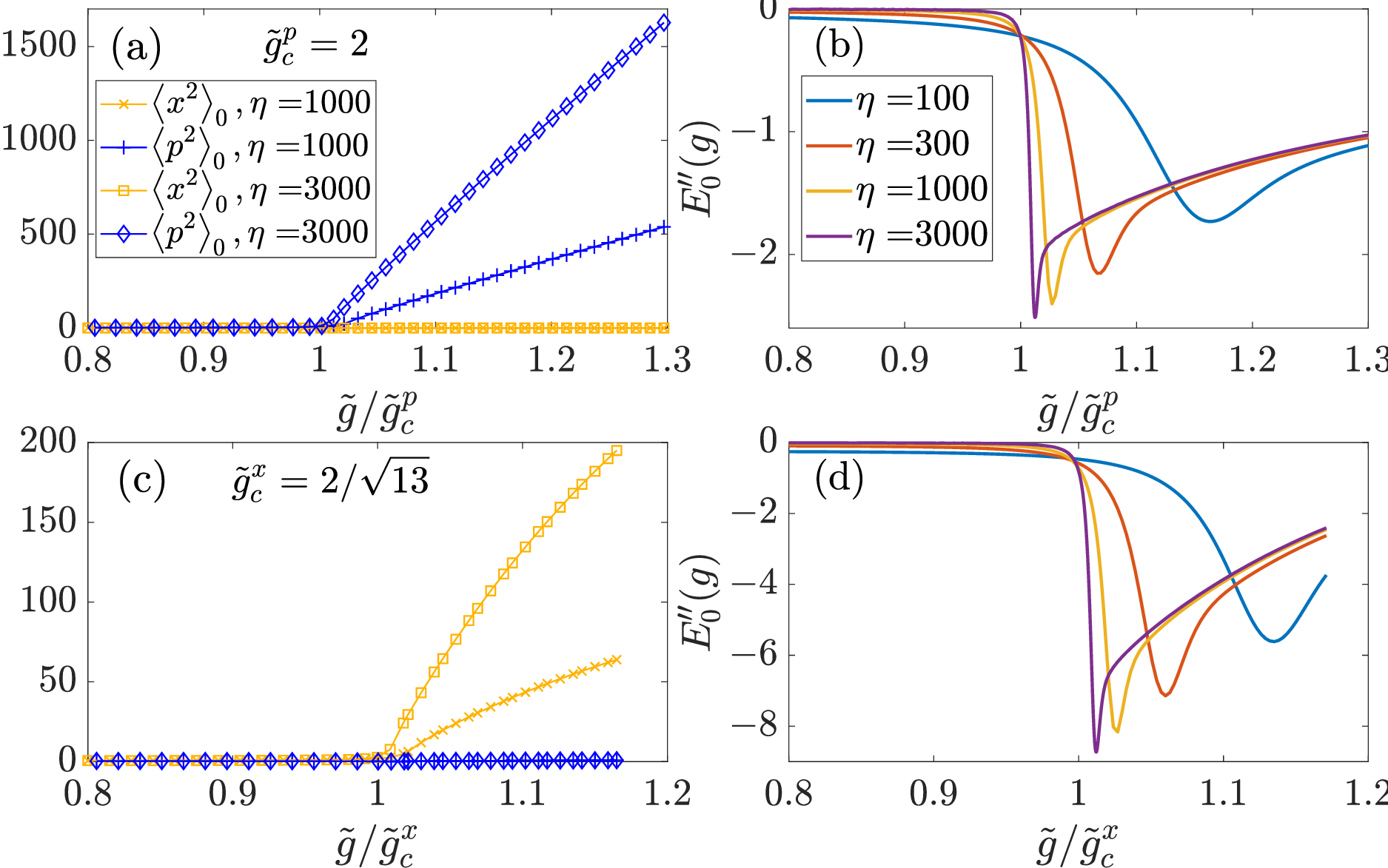}
\caption{Characterization of two types of SRPTs at large effective size $\eta$: (a) and (b) $p$-type for $\tau = 2$, $\kappa = 3$ and (c) and (d) $x$-type for $\tau = 4$, $\kappa = 3$.	(a) and (c) Position-quadrature square $\left\langle x^2 \right\rangle_0$ and the momentum-quadrature square $\left\langle p^2 \right\rangle_0$ of the photon field in the ground state and (b) and (d) second-order derivatives of the ground-state energy.	}
\label{fig_2ndPT}
\end{figure}
To validate the analytical findings, numerical diagonalization is performed on the anisotropic QRM with the $\mathbf{A}^2$ term at a large effective size $\eta$.	Typical parameters are $\tau = 2$ and $ 4$ with $\kappa = 3$, corresponding to a $p$-type SRPT at $\tilde{g}_{c}^{p} = 2$ and an $x$-type SRPT at $\tilde{g}_{c}^{x} = 2 / \sqrt{13}$, as shown in Fig.~\ref{fig_PD}(c).	Figure \ref{fig_2ndPT} presents numerical results for the excitation behavior of the ground-state photonic mode [Fig.~\ref{fig_2ndPT}(a) and ~\ref{fig_2ndPT}(c)] and the second-order derivatives of the ground-state energy $E_{0}^{\prime \prime }(g)$ [Fig.~\ref{fig_2ndPT}(b) and ~\ref{fig_2ndPT}(d)] for various values of $\eta$.	It is evident that the system remains unexcited in the normal phase prior to reaching the critical coupling.	 After the critical coupling of x-type (p-type) SRPT, the position-quadrature square $\left\langle x^{2}\right\rangle _{0}=\left\langle \psi _{0}|x^{2}|\psi_{0}\right\rangle $ (momentum-quadrature square $\left\langle
p^{2}\right\rangle _{0}=\left\langle \psi _{0}|p^{2}|\psi _{0}\right\rangle) $%
of the photon field is excited significantly with increasing $\eta $, while the
other one $\left\langle p^{2}\right\rangle _{0}$ $\left( \left\langle
x^{2}\right\rangle _{0}\right) $ remains unexcited. This behavior clearly indicates the emergence of $x$-type ($p$-type) SRPT.	Further evidence is provided by the sudden drop in $E_0^{\prime\prime}(g)$ around the critical point.	Thus, the second-order nature of both $x$-type and $p$-type SRPTs is convincingly observed numerically, confirming the analytical findings that anisotropy can overcome the no-go theorem.	

\begin{figure}[tbp]
\centering
\includegraphics[width=\linewidth]{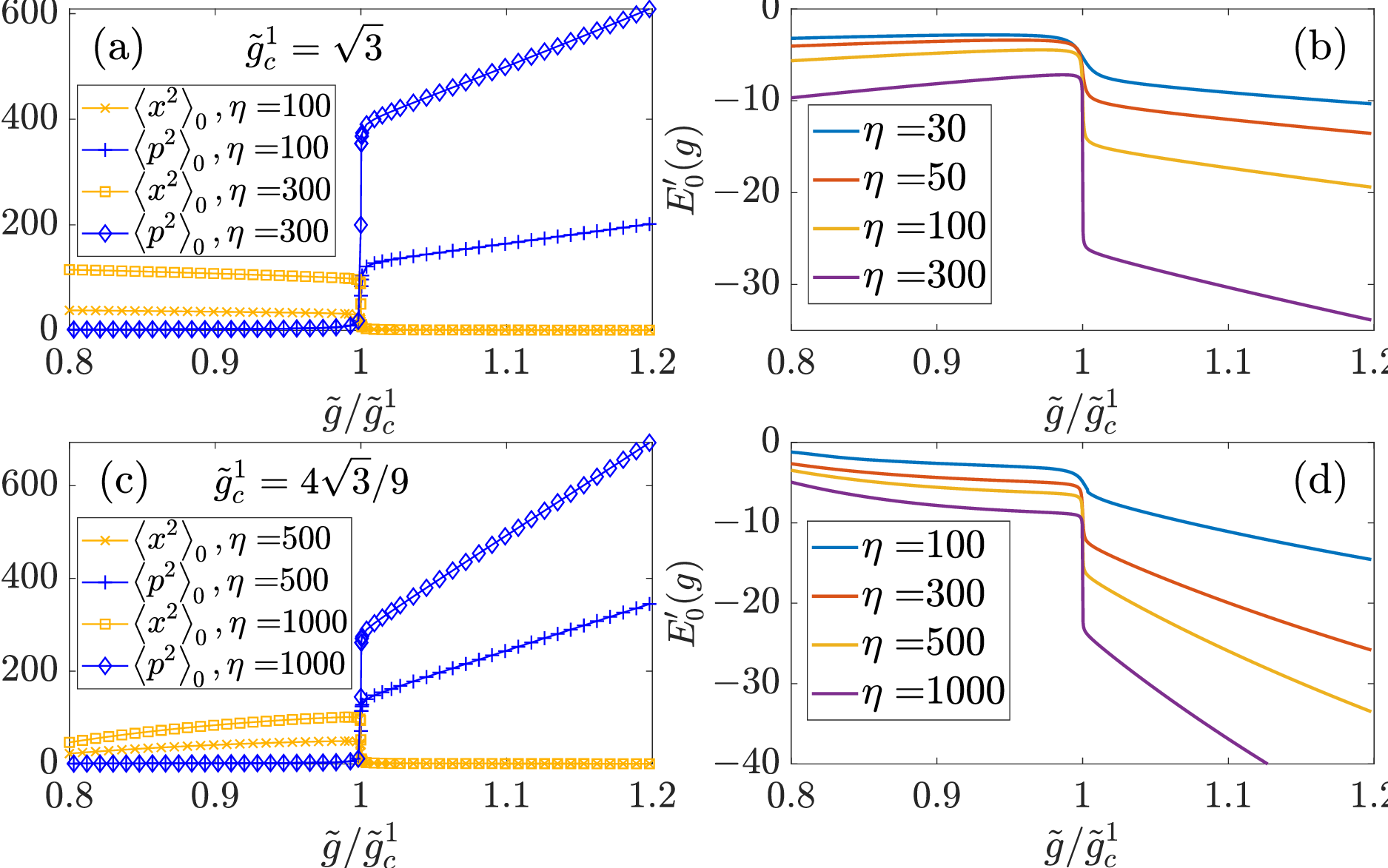}
\caption{Characterization of the first-order phase transition from $x$-type SRP to $p$-type SRP through large effective size $\eta$ calculations for (a) and (b) $\tau = 3$ and $\kappa = 1$ and (c) and (d) $\tau = 4$ and $\kappa = 3$.	(a) and (c) Position-quadrature square $\left\langle x^2 \right\rangle_0$ and momentum-quadrature square $\left\langle p^2 \right\rangle_0$ of the photon field in the ground state and (b) (d) first-order derivatives of the ground-state energy.	}
\label{fig_1stPT}
\end{figure}

We numerically investigate the first-order phase transition from the $x$-type SRP to the $p$-type SRP at the red solid lines corresponding to $\kappa = \kappa_c$ in Fig.~\ref{fig_PD}(b) and \ref{fig_PD}(c).	The field-excitation behaviors and the first-order derivative of the ground-state energy $E_0'(g)$ for $\tau = 3$ and $\kappa = 1$ and for $\tau = 4$ and $\kappa = 3$ are shown in Fig.~\ref{fig_1stPT}, where the corresponding transition coupling constants are $\tilde{g}_c^1 = \sqrt{3}$ and $4\sqrt{3}/9$, respectively.	Before the phase transition, $\left\langle x^2 \right\rangle_0$ is significantly excited, while $\left\langle p^2 \right\rangle_0$ remains unexcited, indicating the $x$-type SRP of the system.	Across the phase transition, the excited field-quadrature square of the photon mode shifts from $\left\langle x^2 \right\rangle_0$ to $\left\langle p^2 \right\rangle_0$.	This indicates a clear transition from the $x$-type SRP to the $p$-type SRP.	Additionally, the first-order derivative of the ground-state energy $E_0'(g)$ drops abruptly at the transition point, showing a clear discontinuous trend with increasing $\eta$.	Therefore, we confirm the first-order nature of the phase transition from the $x$-type SRP to the $p$-type SRP.

In particular, the joint point shown in Fig.~\ref{fig_PD}(c) with $\tau = 3$ and $\tilde{g} = 1$ is of interest, where the phase transition to the $p$-type SRP changes from second order to first order as $\tau$ increases.	Points such as those determined in Eq.~(\ref{gtri}) are surrounded by all three types of quantum phases and are therefore termed the triple point.	The triple point may provide distinguishing critical properties compared to the original SRPT.	To this end, in the following we will extract the critical behaviors of both the standard SRPT and the triple point using finite-size-scaling analysis.	

\begin{figure}[tbp]
\centering
\includegraphics[width=\linewidth]{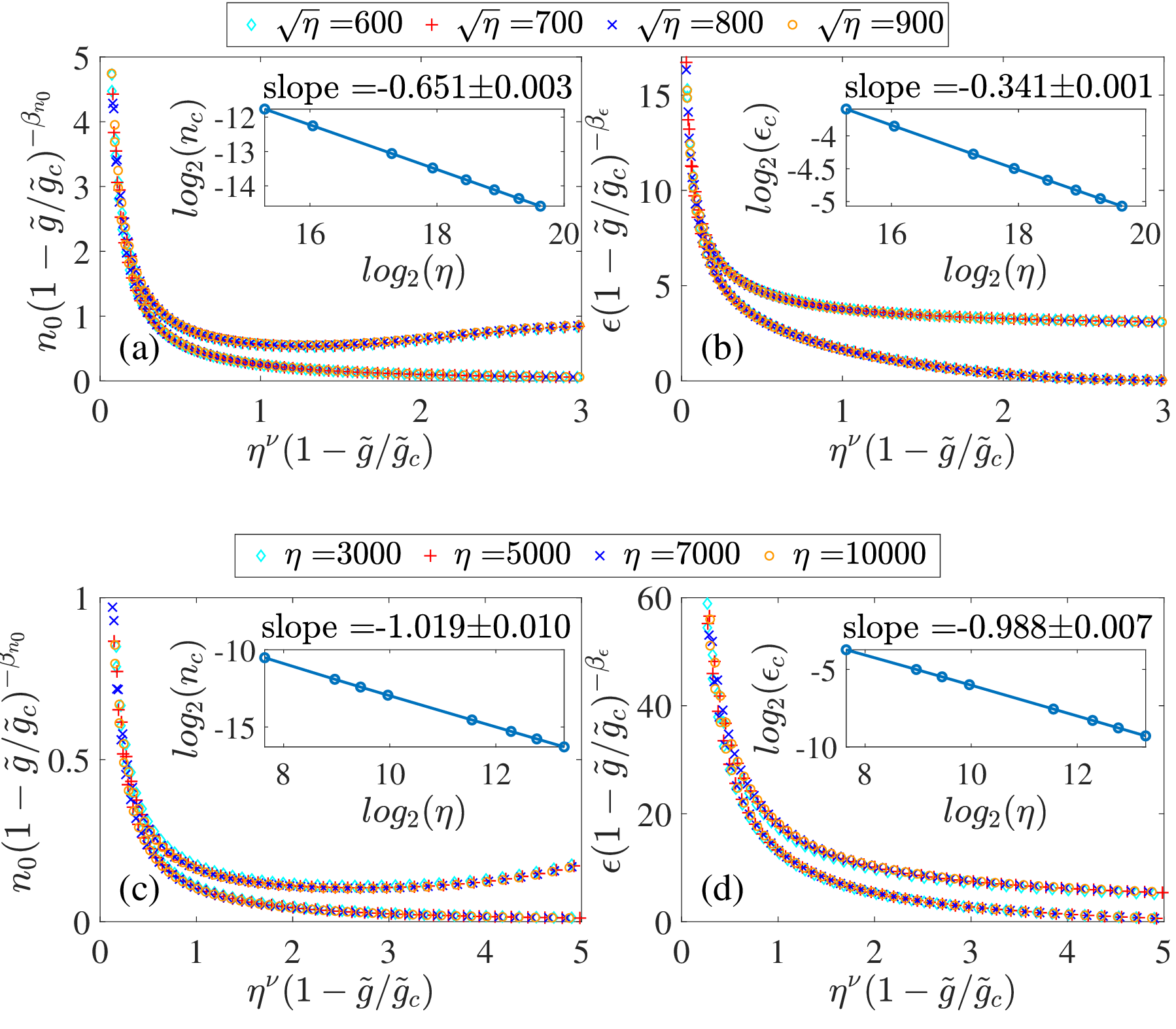}
\caption{Finite-size scaling of (a) and (c) the order parameter $n_{0}=\langle
\protect\psi _{0}|a^{\dagger }a|\protect\psi _{0}\rangle /\protect\eta $
and (b) and (d) the energy gap $\protect\epsilon =E_{1}-E_{0}$ for (a) and (b) 
the original SRPT with $\tau=2$ and $\protect\kappa =3$ and (c) and (d) the triple point with $\tau=3$ and $\protect\kappa =3$ . The insets shows (a) and (c) size $\eta$ dependence of the order parameter $n_{c}$ and (b) and (d)
the energy gap $\protect\epsilon _{c}$ at the corresponding critical point, presented in a log-log scale.}
\label{fig_FSS}
\end{figure}

In the critical regime of a continuous phase transition, the finite-size scaling function of a physical observable $O$ should take the form
\begin{equation*}
O\left( \eta ,\tilde{g}\right) =\left\vert 1-\tilde{g}/\tilde{g}%
_{c}\right\vert ^{\beta _{O}}f\left( \left\vert 1-\tilde{g}/\tilde{g}%
_{c}\right\vert \eta ^{\nu }\right),
\end{equation*}%
where the frequency ratio $\eta$ is the effective size of the system, $\beta_O$ is the critical exponent of the observable $O$, and $\nu$ is the correlation-length critical exponent, which is independent of the observable.	Here we investigate the finite-size scaling behaviors for the order parameter $n_{0}=\left\langle \psi _{0}\left\vert a^{\dagger }a\right\vert \psi_{0}\right\rangle /\eta $ and energy gap $\epsilon =E_{1}-E_{0}$.	The numerical results for the ordinary SRPT at $\tau = 2$ and $\kappa = 3$ are shown in Fig.~\ref{fig_FSS}(a) and ~\ref{fig_FSS}(b).	Note that the curves for different effective sizes $\eta$ collapse well onto a single curve, with the observable-independent correlation length exponent $\nu = \frac{2}{3}$ and the critical exponents $\beta_{n_0} = 1$ ($\beta_\epsilon = \frac{1}{2}$) for the order parameter $n_0$ and energy gap $\epsilon$.	It is not surprising that these critical exponents are the same as those of the DM and the QRM \cite{emary2003prl, vidal2006pra, hwang2015prl, mxliu2017prl, jqyou2018njp}.

The numerical results for the triple point at $\tau=3$ and $\kappa =3$ are shown in
Fig.~\ref{fig_FSS}(c) and ~\ref{fig_FSS}(d). While the  critical exponent
of the order parameter $\beta_{n_{0}}$ remains unchanged, the
correlation-length (energy-gap) critical exponent changes to $\nu ^{T}=1$ ( $%
\beta _{\epsilon }^{T}=1$), which  differs from those in the ordinary
SRPT. Nice power-law behavior of the order parameter and the energy gap at
the critical point ($O_{c}\propto \eta ^{-\gamma _{O}}$) is also demonstrated
in the insets. The fitting value of the scaling exponent $\gamma _{O}$
satisfies the scaling law $\gamma _{O}=\beta _{O}\nu $ very well and further confirms the different correlation-length critical exponent at the multi-critical point. Therefore, it can be concluded that the criticality at the triple point differs from that of the ordinary SRPT.	

Interestingly, the energy gap exponent at the triple point and in the ordinary SRPT can also be derived analytically using an alternative effective Hamiltonian, as provided in the Appendix.	

\section{Conclusion\label{conclusion}}

In this paper we investigated the SRPT in the anisotropic QRM with an arbitrary $\mathbf{A}^2$ term. By performing the Bogoliubov and unitary transformations, we derived a low-energy effective Hamiltonian that allows for SRPTs of both $x$- and $p$-types, as well as a triple point connecting the first- and second-order QPTs.	The second-order SRPT remains robust even when the $\mathbf{A}^2$  term exceeds the value given by the TRK sum rule, as long as the anisotropic qubit-cavity coupling is present, thus surpassing the limitations of the celebrated no-go theorem. In particular, it was revealed that the $x$-type SRP is suppressed as the $\mathbf{A}^2$ term increases, whereas the $p$-type SRP remains robust even for arbitrarily strong $\mathbf{A}^2$ terms.
A rich phase diagram for the anisotropic QRM is derived analytically and confirmed by careful numerical diagonalization at large effective sizes.	Numerical finite-size-scaling analysis revealed that the critical correlation-length exponent and energy-gap exponent $\nu
=1$ and $\beta _{\epsilon }=1$ at the triple point differ from  $\nu
=2/3$ and $\beta _{\epsilon }=1/2$ in the original SRPT. The same energy-gap exponents can also be derived analytically.	

In summary, we theoretically proposed SRPTs in the anisotropic QRM under a strong $\mathbf{A}^2$ term, as dictated by the TRK sum rule. In this model, aside from anisotropy, no other factors, such as qubit-qubit interactions, suppression of the $\mathbf{A}^2$ terms, inter-cavity hopping, periodic driving, or controlled dissipation, are involved. If the proposed model, i.e., an anisotropic qubit-cavity coupling system, is experimentally realized in the deep strong-coupling regime, an SRPT may be observed. To date, the equilibrium SRPT has not been convincingly observed in experiment.	This proposal is one of the most practical candidates for the potential observation of an SRPT and may also provide a new platform for studying quantum critical phenomena.	
	
\textbf{ACKNOWLEDGEMENTS} This work was supported by the National Key R$\&$D  Program of China (Grants No.  2024YFA1408900 and No. 2022YFA1402701),  the National Science Foundation of China (Grants No. 12105001 and No. 12305009), and the Natural Science Foundation of Anhui Province (Grant No. 2108085QA24).	

\textbf{DATA AVAILABILITY}
The data that support the findings of this article are openly available~\cite{data}.
\appendix

\section*{Appendix: Analytical derivation of the energy-gap critical exponent using an alternative effective Hamiltonian	 \label{appendixA}}

In this appendix, we derive an alternative effective Hamiltonian for SRPs. Although it is more complicated than Eq. (\ref{kappa_c}) in the main text, it further provides the energy gap and its critical exponent analytically.	

To achieve this, we apply the following transformations to the Hamiltonian (\ref{Hrscl}):	 $H^{\text{SR}}=U_{{}\sigma _{y}}^{\dagger
}U_{{}\sigma _{z}}^{\dagger }U_{n}^{\dagger }\tilde{H}%
(x+x_{0},p+p_{0})U_{n}U_{{}\sigma _{z}}U_{{}\sigma _{y}}$, where $\tilde{H}%
(x+x_{0},p+p_{0})$ denotes the displacing transformation on $\tilde{H}(x,p)$, $U_{n}=e^{-i(x^{2}+p^{2})\theta _{1}/2}$, $U_{{}\sigma _{z}}=e^{-i\sigma
_{z}\theta _{2}/2}$, and $U_{{}\sigma _{y}}=e^{-i\sigma _{y}\theta _{3}/2}$.
Thus, $H^{\text{SR}}$ is collected as
\begin{eqnarray}  \label{HSR1}
&&H^{\text{SR}}=\frac{x^{2}+p^{2}}{2\eta ^{\prime }}  \notag \\
&&+\sigma _{z}\left[ \cos \theta _{3}+2(G^{+}x_{0}\cos \theta
_{2}-G^{-}p_{0}\sin \theta _{2})\sin \theta _{3}\right] /2  \notag \\
&&+x\sigma _{x}\cos \theta _{3}\left( G_{R}\cos \theta _{-}+G_{cR}\cos \theta _{+}\right)  \notag \\
&&-p\sigma _{y}\left( G_{R}\cos \theta _{-}-G_{cR}\cos\theta
_{+}\right)  \notag \\
&&+C_{\sigma _{x}}\sigma _{x}+C_{\sigma _{y}}\sigma _{y}+C_{x\sigma
_{y}}x\sigma _{y}+C_{p\sigma _{y}}p\sigma _{x}  \notag \\
&&+\frac{x}{\eta ^{\prime }}C_{x}\left|\downarrow \right\rangle \left\langle
\downarrow \right|+\frac{x}{\eta ^{\prime }}|\left\uparrow \right\rangle
\left\langle \right\uparrow |\big[(x_{0}\cos \theta _{1}-p_{0}\sin \theta _{1})
\notag \\
&&\quad \,+\eta ^{\prime }\sin \theta _{3}\left( G_{R}\cos\theta
_{-}+G_{cR}\cos\theta _{+}\right) \big]  \notag \\
&&+\frac{p}{\eta ^{\prime }}C_{y}\left|\downarrow \right\rangle \left\langle
\downarrow \right|+\frac{p}{\eta ^{\prime }}\left|\uparrow \right\rangle
\left\langle \right\uparrow |\big[(x_{0}\sin \theta _{1}+p_{0}\cos \theta _{1})
\notag \\
&&\quad \,+\eta ^{\prime }\sin \theta _{3}\left( G_{R}\sin \theta
_{-}+G_{cR}\sin\theta _{+}\right)\big]  \notag \\
\end{eqnarray}%
where $\left\vert \uparrow \right\rangle $($\left\vert \downarrow
\right\rangle $) is the eigenstate of new Pauli matrix $\sigma _{z}$
satisfying $\sigma _{z}\left\vert \uparrow \right\rangle =\left\vert
\uparrow \right\rangle $ ($\sigma _{z}\left\vert \downarrow \right\rangle
=-\left\vert \downarrow \right\rangle $), $\theta_{\pm} =\theta_{1}\pm\theta_{2}$, $G_{R}=\frac{\tilde{g}^{\prime }}{%
\sqrt{8\eta ^{\prime }}}$, $G_{cR}=\frac{\tilde{g}^{\prime }\tau^{\prime }}{%
\sqrt{8\eta ^{\prime }}}$, $G^{+}=G_{R}+G_{cR}$, $G^{-}=G_{R}-G_{cR}$, and
the coefficients $C_{\sigma _{x}}$, $C_{\sigma _{y}}$, $C_{x\sigma
_{y}}$, $C_{p\sigma _{x}}$, $C_{x}$, and $C_{y}$ are defined as
\begin{eqnarray*}
&&C_{\sigma _{x}}=-\sin \theta _{3}/2+(G^{+}x_{0}\cos \theta
_{2}-G^{-}p_{0}\sin \theta _{2})\cos \theta _{3}, \notag \\
&&C_{\sigma _{y}}=-(G^{+}x_{0}\sin \theta _{2}+G^{-}p_{0}\cos \theta _{2}),
\notag \\
&&C_{x\sigma _{y}}=G_{R}\sin\theta _{-}-G_{cR}\sin\theta _{+},\notag \\
&&C_{p\sigma _{x}}=\cos \theta _{3}\left( G_{R}\sin\theta _{-}
+G_{cR}\sin\theta_{+}\right),\notag \\
&&C_{x}=(x_{0}\cos \theta _{1}-p_{0}\sin \theta _{1})\notag \\
&&\quad \quad \ -\eta ^{\prime }\sin \theta _{3}\left(G_{R}\cos\theta
_{-}+G_{cR}\cos\theta _{+}\right), \notag \\
&&C_{y}=(x_{0}\sin \theta _{1}+p_{0}\cos \theta _{1})  \notag \\
&&\quad \quad \  -\eta ^{\prime }\sin \theta _{3}\left( G_{R}\sin\theta_{-}
+G_{cR}\sin\theta _{+}\right).
\end{eqnarray*}%
It should be noted that if the transformation coefficients satisfy $C_{\sigma _{x}}=C_{\sigma _{y}}=C_{x\sigma _{y}}=C_{x\sigma _{y}}=0$, the fifth line of Eq. (\ref{HSR1}) is eliminated. Moreover, if $C_{x}$ and $%
C_{p} $ vanish, the influence of the last four lines on the ground state can be neglected in the thermodynamic limit. Therefore, when these conditions are satisfied, Hamiltonian (\ref{HSR1}) reduces to the same form as Hamiltonian (\ref{Hrscl}) before the transformations,
\begin{eqnarray}
&\,&\tilde{H}^{\text{SR}}=H^{\text{SR}}\cos \theta _{3}=\frac{x^{2}+p^{2}}{2\eta ^{\prime
\prime }}+\frac{1}{2}\sigma _{z}  \notag  \label{HSR2} \\
&\,&+\frac{\tilde{g}^{\prime \prime }}{\sqrt{8\eta ^{\prime \prime }}}\left[
(1+\tau^{\prime \prime })x\sigma _{x}-(1-\tau^{\prime \prime })p\sigma _{y}\right]
,
\end{eqnarray}%
where the new coefficients are
\begin{eqnarray*}
&&\eta ^{\prime \prime }=\frac{\eta ^{\prime }}{\cos \theta _{3}} \\
&&\tilde{g}^{\prime \prime }=\frac{\tilde{g}^{\prime }\sqrt{\cos \theta _{3}}%
}{2}\big[\cos\theta _{-}(\cos \theta _{3}+1)  \notag \\
&&\qquad \qquad \qquad \quad +\tau^{\prime }\cos \theta _{+}(\cos
\theta _{3}-1)\big],\\
&&\tau^{\prime \prime }=\frac{\cos\theta _{-}(\cos \theta
_{3}-1)+\tau^{\prime }\cos \theta _{+}(\cos \theta _{3}+1)}{\cos
\theta _{-}(\cos \theta _{3}+1)+\tau^{\prime }\cos\theta
_{+}(\cos \theta _{3}-1)}.
\end{eqnarray*}

Next, performing a similar unitary transformation $H_{\text{eff}}^{\text{SR}}=e^{-S}%
\tilde{H}^{\text{SR}}e^{S}$ with the generator $S=-i\tilde{g}^{\prime \prime }\left[
(1+\tau^{\prime \prime })x\sigma _{y}+(1-\tau^{\prime \prime })p\sigma _{x}\right]
/\sqrt{8\eta ^{\prime \prime }}$, the effective low-energy Hamiltonian for
the SRP is obtained as
\begin{equation}
H_{\text{eff}}^{\text{SR}}\ \simeq \frac{x^{2}+p^{2}}{2\eta ^{\prime
\prime }}-\frac{{{}\tilde{g}^{\prime \prime }}^{2}}{8\eta ^{\prime \prime }}%
\left[ (1+\tau^{\prime \prime })^{2}x^{2}+(1-\tau^{\prime \prime })^{2}p^{2}\right]
.  \label{Heff_SR}
\end{equation}%
The energy gap thus is
\begin{equation*}
\epsilon =\frac{\gamma \omega }{4}\sqrt{(4-{{}C_{\epsilon }^{x}}^{2}\cos
^{3}\theta _{3})(4-{{}C_{\epsilon }^{p}}^{2}\cos \theta _{3})},
\end{equation*}%
with the coefficients $C_{\epsilon }^{x}=\tilde{g}^{\prime }\left( \cos \theta _{-}+\tau^{\prime }\cos \theta _{+}\right) $ and $C_{\epsilon
}^{p}=\tilde{g}^{\prime }\left( \cos\theta _{-}-\tau^{\prime
}\cos\theta _{+}\right) $.

To determine the effective Hamiltonian explicitly, we need to combine it with the solution of the condition $C_{\sigma_x}=C_{\sigma_y}=C_{x%
\sigma_y}=C_{p\sigma_x}=C_{x}=C_{y}=0$. There are two possible combinations of the transformations.

First, we set the transformation coefficients as $\theta _{1}=\theta
_{2}=0,p_{0}=0$, and
\begin{eqnarray}
\theta _{3} &=&\arccos {\frac{4\gamma ^{2}}{{\tilde{g}}^{2}(1+\tau)^{2}}},
\notag \\
x_{0} &=&\sqrt{\frac{\eta }{8\gamma ^{3}}}\tilde{g}(1+\tau)\sin \theta _{3}.
\end{eqnarray}%
In this scenario, following the displacement transformation $%
x\rightarrow x+x_{0}$, the bound oscillator-form effective
Hamiltonian implies a ground state with  $\langle x^{2}\rangle \propto
x_{0}^{2}\propto \eta $ and $\langle p^{2}\rangle \propto \eta ^{0}$. This characterizes the $x$-type SRP nature of the superradiant ground state. Furthermore, the energy gap can be expressed as
\begin{eqnarray}
&\,&\epsilon _{\text{xSRP}}=\frac{2\gamma \omega }{(1+\tau)^{3}\tilde{g}_{c}^{x}%
\tilde{g}^{2}}  \label{epsln_x} \\
&\,&\times \sqrt{\left[ \tilde{g}^{2}(1+\tau)^{2}+4\gamma ^{2}\right] (\tilde{g}%
^{2}-{{}\tilde{g}_{c}^{x}}^{2})\left[ 4\tau-(1-\tau)^{2}\kappa \tilde{g}^{2}\right]
}.  \notag
\end{eqnarray}%

Second, we can also set the transformation coefficients as	 $\theta
_{1}=\theta _{2}=\pi /2,x_{0}=0$, and
\begin{eqnarray}
\theta _{3} &=&\arccos {\frac{4}{{\tilde{g}}^{2}(1-\tau)^{2}}}  \notag \\
p_{0} &=&-\sqrt{\frac{\eta }{8\gamma }}\tilde{g}(1-\tau)\sin \theta _{3}.
\end{eqnarray}%
In the scenario, considering the displacement transformation $p\rightarrow
p+p_{0}$ performed, the bound oscillator-form effective Hamiltonian
implies a ground state with $\langle x^{2}\rangle \propto \eta ^{0}$
and $\langle p^{2}\rangle \propto p_{0}^{2}\propto \eta $. This characterizes the $p$-type SRP nature of the superradiant ground state and yields the energy gap as
\begin{eqnarray}
&&\epsilon _{\text{pSRP}}=\frac{2\omega }{(1-\tau)^{3}\tilde{g}_{c}^{p}\tilde{%
g}^{2}}  \label{epsln_p} \\
&&\times \sqrt{\left[ \tilde{g}^{2}(1-\tau)^{2}+4\right] (\tilde{g}^{2}-{{}%
\tilde{g}_{c}^{p}}^{2})\left[ (1-\tau)^{2}\kappa \tilde{g}^{2}-4\tau\right] }.
\notag
\end{eqnarray}%
Next, we analyze the critical behavior of the energy gap. Along with the oscillator nature of the normal-phase system, as revealed by the effective Hamiltonian (\ref{Heff}), the energy gap  $\epsilon $ in the normal phase is given by
\begin{equation}
\epsilon =\frac{\omega }{4}\sqrt{\{4-\left[ (1+\tau)^{2}-4\kappa \right] \tilde{%
g}^{2}\}\left[ 4-(1-\tau)^{2}\tilde{g}^{2}\right] }  \label{epsln_np}
\end{equation}%

When approaching the triple point at $\tau=\kappa $, $\tilde{g}%
_{c}^{x} $ and $\tilde{g}_{c}^{p}$ should coalesce to $\tilde{g}_{c}^\text{tri}$.  In this case, both the first and second radical factors of the normal-phase energy gap in Eq.~(\ref{epsln_np}) tends to vanish, leading to a distinct critical
behavior. Specifically, when approaching $\tilde{g}_{c}^\text{tri}$ with fixed
and identical $\tau$ and $\kappa $ from the normal phase, the energy gap
reduces to $\epsilon =\omega (1-\tilde{g}^{2}/{{}\tilde{g}_{c}^\text{tri}}^{2})$
and vanishes as $\epsilon \propto (\tilde{g}_{c}^\text{tri}-\tilde{g})$.
Similarly, when approaching $\tilde{g}_{c}^\text{tri}$ in the same way, but from the SRP, Eq.~(\ref{epsln_p}) predicts
the vanishing of the energy gap as
\begin{equation*}
\epsilon _{\text{pSRP}}\propto \sqrt{(\tilde{g}-\tilde{g}_{c}^{p})\left(
\tilde{g}-2\sqrt{\tau/\kappa }/\left|1-\tau\right|\right) }=\tilde{g}-\tilde{g}_{c}^\text{tri}.
\end{equation*}%
Thus, this analysis supports a distinct energy-gap critical exponent $\beta
_{\epsilon }=1$ at the triple point, compared to $\beta _{\epsilon }=1/2$
for the  original SRPT, which is consistent with the numerical finite-size-scaling analysis in the main text.

\end{document}